\documentclass[oneside,english,sigconf,sigconf]{acmart}
\usepackage[T1]{fontenc}
\usepackage[latin9]{inputenc}
\usepackage{amsthm}
\usepackage{graphicx}

\makeatletter
\numberwithin{equation}{section}
\numberwithin{figure}{section}

\makeatother

\usepackage{babel}
\begin{document}

\title[Analyzing the HCP Datasets using GPUs]{Analyzing the Human
Connectome Project Datasets using GPUs}

\subtitle{The Anatomy of a Science Engagement}

\author{John-Paul Robinson}
  \affiliation{%
  \department{Research Computing}
  \institution{University of Alabama at Birmingham}
  \city{Birmingham}
  \state{Alabama}
  \postcode{35294}}
\email{jpr@uab.edu}

\author{Thomas Anthony}
  \affiliation{%
  \department{Research Computing}
  \institution{University of Alabama at Birmingham}
  \city{Birmingham}
  \state{Alabama}
  \postcode{35294}}
\email{tanthony@uab.edu}

\author{Ravi Tripathi}
  \affiliation{%
  \department{Research Computing}
  \institution{University of Alabama at Birmingham}
  \city{Birmingham}
  \state{Alabama}
  \postcode{35294}}
\email{ravi89@uab.edu}

\author{Sara A. Sims}
\affiliation{%
  \department{Psychology}
  \institution{University of Alabama at Birmingham}
  \city{Birmingham}
  \state{Alabama}
  \postcode{35294}}
\email{snolin@uab.edu}

\author{Kristina M. Visscher}
\affiliation{%
  \department{Neurobiology}
  \institution{University of Alabama at Birmingham}
  \city{Birmingham}
  \state{Alabama}
  \postcode{35294}}
\email{kmv@uab.edu}

\author{Purushotham V. Bangalore}
\affiliation{%
  \department[0]{Computer Science}
  \department[1]{Research Computing}
  \institution{University of Alabama at Birmingham}
  \city{Birmingham}
  \state{Alabama}
  \postcode{35294}}
\email{puri@uab.edu}

\keywords{Slurm batch scheduler, high performance computing, GPU, research support}

\begin{abstract}
This paper documents the experience improving the performance of a
data processing workflow for analysis of the Human Connectome Project's
HCP900 data set. It describes how network and compute bottlenecks
were discovered and resolved during the course of a science engagement.
A series of computational enhancements to the stock FSL BedpostX workflow
are described. These enhancements migrated the workflow from a slow
serial execution of computations resulting from Slurm scheduler incompatibilities
to eventual execution on GPU resources, going from a 21-day execution
on a single CPU core to a 2 hour execution on a GPU. This workflow
contributed a vital use-case to the build-out of the campus compute
cluster with additional GPUs and resulted in enhancements to network
bandwidth. It also shares insights on potential improvements to distribution
of scientific software to avoid stagnation in site-specific deployment
decisions. The discussion highlights the advantages of open licenses
and popular code collaboration sites like GitHub.com in feeding contributions
upstream.
\end{abstract}

\maketitle
\renewcommand{\shortauthors}{Robinson and Bangalore}

\section{Introduction}

Research support engagements to facilitate use of advanced computing
resources are crucial to improve scientific outcomes and ensure efficient
use and operation of resources. The UAB Visual Brain Core (now Civitan
International Neuroimaging Laboratory (CINL) Brain Core) is a collaboration
between neuro-imaging labs and the research computing support group
in central IT~\cite{Core2017(accessedMarch2018)}. It was created
to improve the availability of compute and software resources across
campus neuro-imaging research groups. Its mission is to help produce
high quality, cutting edge research examining the visual brain. To
achieve this, the core helps investigators overcome some common barriers
to performing high quality vision research, and provides forums to
discuss new ideas and research techniques. The core offers weekly
office hours to assist researcher access to computing resources.

Campus compute cluster support for high-performance and high-throughput
computing is provided by the research computing support group that
is part of the central IT organization. The group operates a recently
acquired 2800 core compute cluster with six petabytes of GPFS storage
and 72 P100 GPUs. This cluster was built over the past three years
and represents a commitment to continued investment in computing support
for the campus research community~\cite{Bakken2017(AccessedMarch2018)}.

Opportunities for science engagement often occur serendipitously through
ordinary use of resources. The engagement described in this paper
resulted from efforts by researchers to prepare for participation
in a distributed conference series focused on investigations of the
the brain. The investigation sought to focus on the HCP900 data set
released by the Human Connectome Project~(HCP)~\cite{Essen2012,Glasser2013}.

\section{Brainhack Global at UAB}

Brainhack Global is a conference series inspired by the experience
of software hackathons~\cite{Brainhack2017(accessedMarch132018)}.
Similar to hackathons in the tech sector, much of the schedule is
left open for attendees to work together on projects of their choosing.
Brainhacks are not \textquotedblleft coding sprints\textquotedblright{}
or exclusive to programmers, but rather are open to brain scientists
from all backgrounds. Part of the goal is to get people with different
backgrounds working together to understand the brain. Brainhack seeks
to convene a global community of researchers across disciplines from
across the globe to work on innovative projects related to neuroscience.
Each site sponsors a local Brainhack event and shares its activities
with the global community. UAB was a participating site in the Brainhack
Global 2017 event held March 2-5, 2017~\cite{UAB-Brainhack2017(accessedMarch2018)}. 

As part of this event one of the participating researchers set out
to prepare the HCP900 dataset for analysis \textquotedblleft in a
few days\textquotedblright{} on the campus 2300-core compute cluster.
The HPC900 is provided by the Human Connectome Project (HCP) which
aims to provide an unparalleled compilation of neural data, an interface
to graphically navigate this data, and the opportunity to achieve
never before realized conclusions about the living human brain. With
the help of the HCP, it is possible to imagine navigating the brain
in a way that was never before possible: flying through major brain
pathways, comparing essential circuits, zooming into a region to explore
the cells that comprise it, and the functions that depend on it.

The HCP900 included structural and diffusion data on 727 subjects,
aged 22-36, 44\% male. The goal was to facilitate use of FSL\textquoteright s
Probtrackx for probablisitic tractography from both central and far-peripheral
regions of interest. This required preprocessing the images to provide
the data set inputs needed by the Probtrackx~\cite{Behrens2007}.
This preprocessing was accomplished with the Bedpostx tool in FSL.

\section{Scaling Network Bandwidth}

\begin{figure}
\includegraphics[scale=0.25]{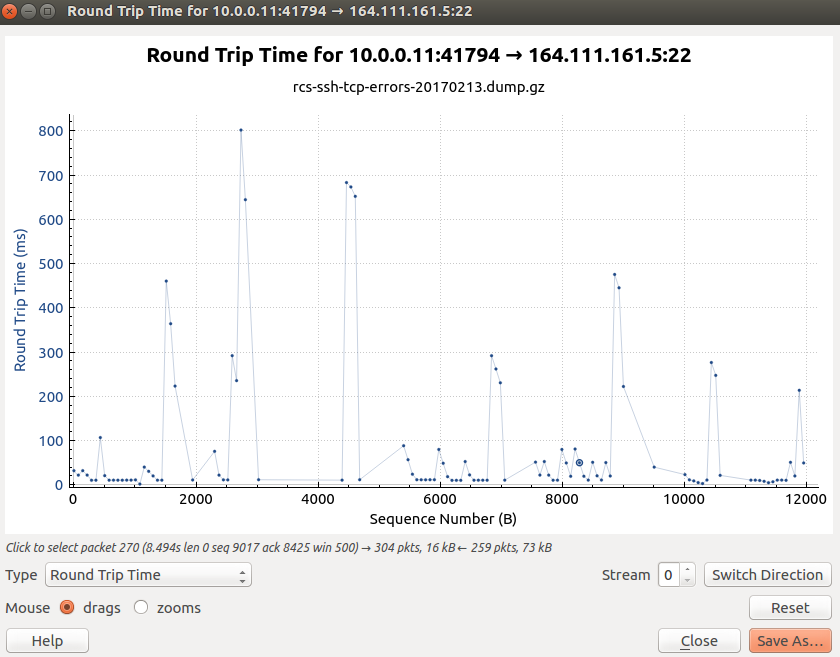}

\caption{\label{fig:Wireshark-TCP-Latency}Wireshark TCP Latency Plot for SSH
connection}
\end{figure}

The Human Connectome Project makes multi-Terabyte data sets available
by way of the Aspera data transfer client which provides high throughput
data transfers via managed UDP data flows~\cite{aspera}. The research
support team became aware of the download of the HCP900 data set by
way of reports from users who were experiencing connectivity issues
to cluster VNC sessions and latency issues in their interactive SSH
sessions to the cluster login node. After investigating the network
layer performance and capturing SSH flows to the login node, we confirmed
that the interactive keystroke latencies were directly measurable
on the TCP packet flow via a Wireshark plot, see Figure~\ref{fig:Wireshark-TCP-Latency}.
The university network team that operates the Intrusion Protection
System (IPS) device in front of the cluster identified an on-going
UDP transfer as a potential cause. Interrupting the Aspera data transfer
resolved all interactive latency issues and confirmed the source of
the network contention was the download of the HCP900 data set. Inspecting
the Aspera client configuration revealed that the client is set to
limit transfers to 1Gbps. It was difficult to understand, however,
that these flows would be causing issues on the 10Gbps rated network
interface of the cluster login node. 

The university network was undergoing improvements and upgrades to
bandwidth facilitated by a campus Science DMZ grant~\cite{UAB-IT2015(accessedMarch2018)}.
This grant provided equipment for 10Gbps links to shared research
resources and motivated additional campus investment across the network~\cite{UAB-IT2016(accessedMarch2018)}.
HPC cluster components are connected internally via 10Gbps or better
Ethernet links. Further investigation of the campus network path to
the cluster revealed that an inter-switch connection from the cluster
network fabric to the campus was still running at 1Gbps. This network
link was clearly constrained and now explained why the default Aspera
configuration was overloading what should have been adequate bandwidth
on the cluster login node. The Aspera client was reconfigured to approximately
70\% of its default bandwidth to allow the transfer to continue without
impacting other users.

\section{Scaling Compute Throughput}

The Oxford Centre for Functional MRI of the Brain (FMRIB) produces
the FMRIB Software Library (FSL), a library of tools to process FMRI,
MRI, and DTI images of the brain~\cite{Behrens2003,Jenkinson2012}.
These tools and their associated pipelines are frequently used by
imaging groups on campus to process research subject data sets. The
FMRIB Diffusion Toolbox (FDT) library provides utilities for processing
diffusion weighted MRIs. Bedpostx is a key preprocessing component
of this toolbox that generates probabilities of crossing fibers within
a voxel~\cite{Behrens2003,Behrens2007}. This provides the necessary
inputs for modeling nerve fiber pathways in the brain via tractography.
This is a computationally intensive process that must analyze each
voxel of an MRI. Bedpostx makes it easy for a user to identify a collection
of MRIs for processing and takes care of managing the execution details
for stepping through each subject MRI and producing the necessary
output files.

\subsection{Software Tools on the Cluster}

Software application modules are provided to users of the campus compute
cluster so that they can easily activate applications in their environment.
The modules customize their execution path and other environment settings
needed to run an application. The software and modules are deployed
using Easybuild, a software packaging and deployment suite that simplifies
the effort for research support groups to provide software packages
across a wide variety of science domains~\cite{Geimer2014}. Easybuild
includes packages for building FSL, making it easy to deploy this
tool for the campus neuro-imaging community.

A typical neuro-imaging workflow will involve the user activating
their FSL module and then running the Bedpostx application with a
provided input directory of images. In a non-batch scheduling environment,
Bedpostx processes each subject MRI serially on a single core. Bedpostx
is actually a collection of shell scripts that feeds input to the
tool xfibres. If the script detects a Sun Grid Engine (SGE) batch
scheduling environment (by way of the SGE\_ROOT environment variable),
it will arrange to split each MRI into its constituent slices and
submit each slice as a job to the batch scheduler. This will significantly
improve the processing time by reducing the total time per subject
to the time it takes to process the slice containing the most voxels,
given enough cores to run all slices simultaneously. Images in the
HCP900 data set contained approximately 145 slices.

\subsection{Understanding FSL Application Defaults}

Research deadlines put pressure on completing data processing steps
as quickly as possible. The research team was working to prepare the
HCP900 data set for analysis in time for their upcoming Brainhack
event. Producing connectivity maps for each subject was desired. With
FSL readily available on the cluster, the team set out to pre-process
the complete data set. After some preliminary tests, it was observed
that Bedpostx only processed one image at a time and was taking an
extensive amount of time to complete. Based on prior experience with
FSL and the campus batch scheduler, the team produced a job submit
script that would submit each subject individually. The jobs were
configured to request 6 cores with 14G per-core (84G total per job)
in the longest job runtime configuration (6 days and 6 hours). The
hope of this resource request was that it would improve the throughput
on each job and provide enough time to complete the analysis. Due
to the time constraint, 800+ jobs were submitted at once. Due to limitations
in the scheduler configuration that allowed one user to consume as
many resources are requested, this job configuration quickly overwhelmed
the resources available on the cluster. This resulted in all cores
on the cluster being consumed by this single workflow. Naturally other
users objected to the resource starvation for their own workflows.

This cluster event caused the research support group to engage in
the effort. Conversations with the research team lead to an understanding
of the goals and the existing assumptions about FSL. The support team
investigated the team's job scripts and researched the operational
assumptions of FSL and Bedpostx. The most significant finding was
that Bedpostx was only programmed to detect SGE environments. As a
result of the recent campus compute upgrades and modernization of
batch scheduling capabilities, the campus cluster environment had
been migrated from the long-standing SGE batch scheduler to the Slurm
batch scheduler four months prior. Because of this, Bedpostx failed
to detect a batch computing environment. This caused it to analyze
each MRI on a single-core. This was counter to the researcher's prior
experience with Bedpostx, wherein it would automatically run efficiently
via the (then SGE) batch scheduler. Finally, the resources requested
for each job did little to improve performance because only one of
six requested cores was actually performing the computation.

\subsection{Adapting to the Local Environment}

After some research on Bedpostx and reading the code, it was clear
that Bedpostx acts as a wrapper script for another script fsl\_sub,
which actually submits brain slices for processing. We also realized,
based on reading the code for fsl\_sub, that it was designed exclusively
for the SGE scheduler. Since the batch computing environment was now
using the Slurm scheduler, we needed to modify fsl\_sub to recognize
our Slurm environment and submit the brain image slices to it in parallel.
This modification was facilitated by an existing FSL patch provided
by Trinity College Dublin~\cite{Research-IT2015(accessedMarch2018)}.

\begin{figure*}[t]
\includegraphics[scale=0.3]{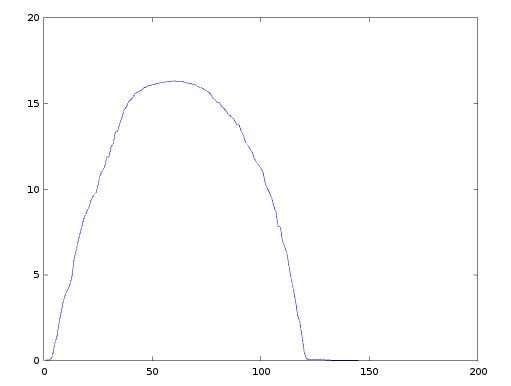}\includegraphics[scale=0.3]{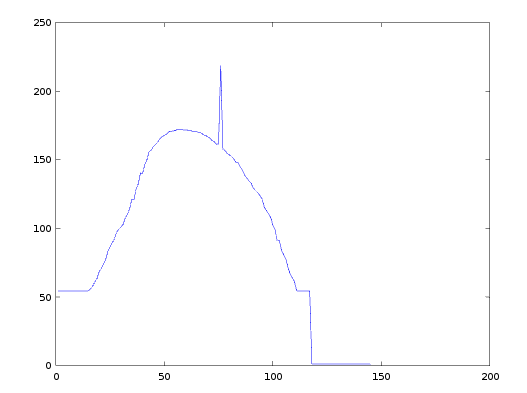}\includegraphics[scale=0.3]{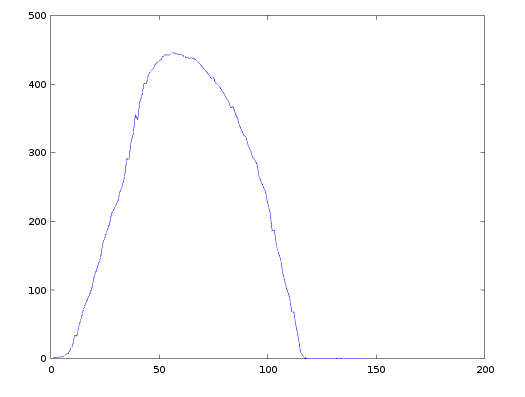}

\caption{\label{fig:xfibres-file-size}xfibres run: file size (MB), RAM used
(MB), and processing time (seconds) per-slice in a single-subject
MRI}

\end{figure*}

We created a test copy of FSL software based on the Trinity patch.
Development was tracked using the research support teams GitLab environment.
Preliminary changes were made to adjust for local partition naming
and job request parameters. With batch submission working, a series
of test runs were initiated to observe performance. The updated fsl\_sub
executable was now able to run all the brain image slices in parallel
when using Slurm batch scheduling. The performance of each slice was
clearly correlated to the number of voxels per-slice which contain
actual brain imaging data (as opposed to a non-data mask). The more
data in the slice, the more RAM required and the longer the compute
time, as seen in~Figure~\ref{fig:xfibres-file-size}. Given that
each slice is processed independently this kind of scaling provides
a solid indication that additional slicing of the data can further
improve performance.

This analysis allowed the research support group to tune the requested
parameters in the fsl\_sub to match the data set, maximizing resource
efficiency across jobs. The maximum run time of an individual slice
was chosen as the default job request time and job dependencies were
introduced to allow the MRI pre- and post-processing (slice decomposition
and reconstruction) to operate in a dependency workflow. With Slurm
support in place, the total time for processing of a single subject
was reduced from 21 days to 7 hours when running all the 145 brain
image slices in parallel on 145 cores/cpus. Nonetheless, with 800+
images to process from the HCP900 collection, the entire processing
time would have still taking around 21 days to process on the campus
cluster, even given all available cores (2300 then) to this single
workflow.

\subsection{Maximizing Throughput with GPU Computing}

CPUs use chip space for large, fast memory caches to speed execution
of serial codes. GPUs use chip space to add 1000\textquoteright s
of cores to speed execution of data-parallel codes. The GPU is especially
well-suited to address problems that can be expressed as data-parallel
computations - the same instructions are executed on many data elements
in parallel. Many applications that process large data sets can use
a data-parallel programming model to speed up their computations.
The analysis of xfibres suggested this was very amenable to GPU computing
and, indeed, it was discovered that FSL included a version of xfibres
to run on NVIDIA GPUs with a documented 200x acceleration over a singe
CPU core~\cite{Hernandez2013}.

The recent campus cluster expansion included four nodes with a K80
GPU to support exploration with GPU computing. In parallel to the
Bedpostx customization for Slurm, the research support team tested
the xfibres\_gpu application provided with FSL using the CUDA Toolkit
v7.5.18. The xfibres\_gpu code does a simple cudaGetDevice() as its
only initialization step to get the current GPU device to run its
commands.txt~\cite{NVIDIA2017(accessedMarch132017)}. The commands.txt
file is automatically generated and lists the application and parameter
combination needed to process the data set.

A custom job script was developed to directly submit xfibre\_gpu runs
to the K80 GPU nodes. Additional changes to Bedpostx would have been
required to detect GPU capabilities within the Slurm environment.
It was easier to simply process each MRI as a single job on the GPU.
Each job needed 8 GB of RAM. Each subject was now able to complete
in slightly over 2 hours (2 hours 8 minutes). These runs were able
to match the documented speedup compared to the original single-core
processing time of 21 days.  ((21 days {*} 24 hours) / 2 hours 8 minutes
= 236x speedup). An additional benefit was that the image processing
could now be moved to dedicated nodes freeing the rest of the cluster
to serve other users. Unfortunately, due to the large number of subjects
in the HCP900 data set and having only four GPU nodes, processing
the entire data set still required 21 days.

\section{Results and Outcomes}

\subsection{Network Enhancements and the Science DMZ}

In response to the discovery of the 1Gbps constriction point, the
network team prioritized the upgrade of the problematic 1Gbps link
to 10Gbps. This result was obtained within a few days of discovery.
Since this upgrade, no similar bandwidth competition issues have been
reported. Further testing and use of other transfer protocols, however,
has shown the next constriction point is the IPS itself. This motivates
moving all significant data flows, including Aspera facilitated transfers,
to the Science DMZ fabric being constructed for the cluster.

The ability to run network performance tests that follow the path
of user data is crucial to determining performance expectations for
data transfers. Regular testing helps identify bandwidth availability
patterns and can alert when performance falls outside the norm. The
PerfSONAR infrastructure that is part of the ESNet Science DMZ initiative
provides valuable tooling for these efforts~\cite{Hanemann2005,Dart2013}.
The campus has built a Science DMZ as part of the NFS grant with two
data transfer nodes (DTNs) and a PerfSONAR node, each with 40Gbps
connectivity to 100Gbps network that feeds the campus.

Aspera transfers have been tested on this network and max out at slightly
above 1Gbps. This suggests any further transfer improvements will
depend on improvements to the HCP900 data delivery network. Nonetheless,
adding performance testing software to the cluster login nodes, where
most transfers still occur, would have revealed the 1Gbps constriction
immediately. This configuration has been problematic because cluster
login nodes sit on a private network behind a firewall. This configuration
complicates network bandwidth tests because some end-to-end tests
only advertise the internal network address of the testing node. Installing
network performance testing tools on all relevant data endpoints is
a priority for the cluster support team. The DTNs are also in the
process of being connected directly to cluster storage which will
make them available to production data transfers.

\subsection{Local Software Forks and Feeding Improvements Upstream}

Maintaining local modifications to software suites across releases
requires considerable effort. This is complicated by licensing that
limits the ability to distribute modifications of the FSL distribution.
Licensing choices can be used to shape the nature of the downstream
user community. The changes to FSL described in this paper must be
maintained locally. While the Trinity patch formed a useful foundation
and sped adaption, site-specific changes were still necessary. In
mature open-source software, such site changes are often found in
configuration files. The upstream provider has little motivation to
develop enhancements that don't serve their local needs, but a vibrant
community around an open source software product can enable downstream
users to build these features into the software. This type of community
is readily facilitated by popular code collaboration sites like GitHub.com.
The changes to the FSL submit functionality to support Slurm have
been published on GitHub.com~\cite{Robinson2018(accessedJune2018)}.

\subsection{Expanding Impact of GPU Computing}

\begin{figure}
\includegraphics[scale=0.2]{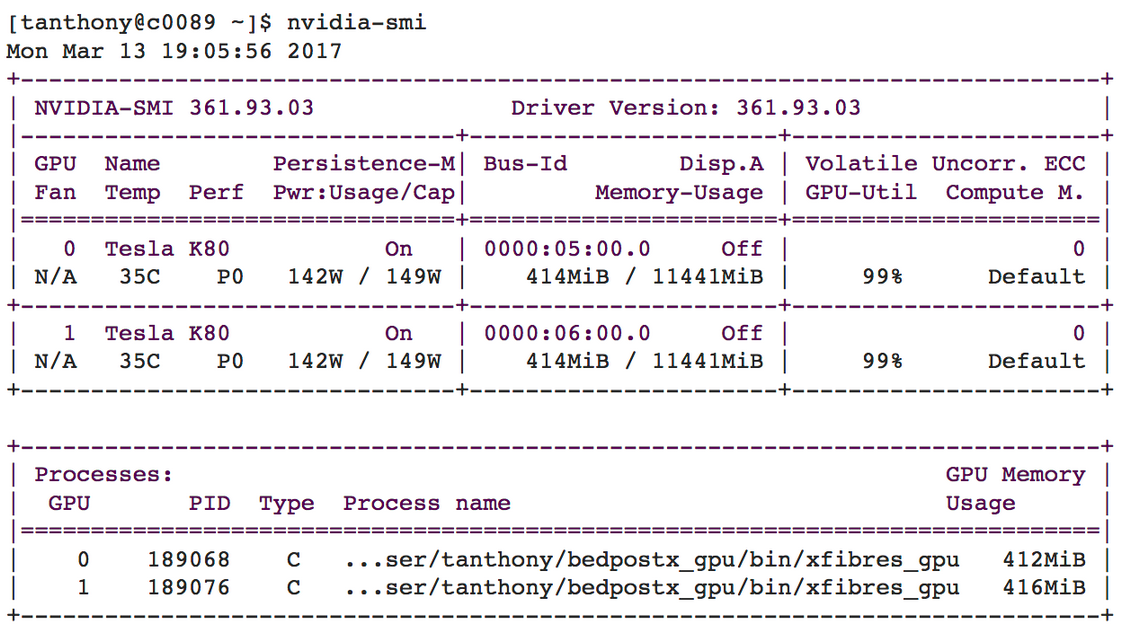}

\caption{\label{fig:Detection-of-NVIDIA}Detection of NVIDIA GPUs via CUDA
environment}
\end{figure}

The potential of GPU computing to significantly increase the throughput
of computationally intense work loads and simultaneously reduce the
data center footprint makes it an attractive option for investments
focused on expanding computing at the campus. The research support
group used the Bedpostx workflow as a test case to evaluate its acquisition
decisions for compute expansion. Detecting the number of GPU devices
on the system can enhance the throughput further by splitting the
slice collection across each GPU or assigning one whole subject per
GPU. In the particular case of the site's existing NVIDIA K80 GPU
nodes, each K80 can technically be treated as two K40 GPUs. This enables
two xfibers\_gpu processes to run on each half of the K80 GPU, doubling
the per-subject throughput of a single GPU node as shown in~Figure~\ref{fig:Detection-of-NVIDIA}.
This is very useful for multi-GPU systems. Though we could not complete
the automation of this splitting process before the Brainhack event,
we were able to manually test and demonstrate this on our existing
NVIDIA K80 nodes and vendor provided test system with 4x NVIDIA P100
GPUs. A single P100 GPU was demonstrated as capable of processing
a single subject in 30 minutes, suggesting a four GPU cluster node
has the potential to process a single subject in less than 10 minutes
with a four-way split to the MRI image. This use case contributed
to the decision to acquire 18 compute nodes with four P100 GPU nodes
six months later. This expanded compute fabric has the potential to
process the HCP900 dataset discussed here in less than 24 hours (tested
at 40 seconds of compute time on a single subject across 68 GPUs).
While this pre-processing step is not run often, the significance
of moving a big data workflow from 21 days per subject to less than
1 day for the entire dataset can change how researchers envision their
science and what can be accomplished.

\subsection{Improvements to Batch Scheduling}

Several improvements to the Slurm scheduling environment were implemented
in the months following the Brainhack. The improvements focused on
introducing multi-factor priority scheduling to implement a fair share
policy for access to resources. The new policy provides a priority
boost to users who have not used the cluster over the past month.
This helps ensure resource availability for less frequent cluster
users. Also quality of service constraints have been introduced to
prevent a single user for consuming more than an agreed on fraction
of cluster compute and memory resources, currently approximately 15\%
and 30\%, respectively. This prevents the type of job flooding experienced
during the HCP900 Bedpostx analysis and provides a mechanism for the
operations team to adjust capacity as needed and in a controlled fashion
for select work loads. 

It may be seen as an omission to not have balanced usage policies
in place when migrating to a new scheduler. The depth of understanding
needed to implement all desired policies often exceeds the experience
level and project time lines of small campus support teams. Implementing
these policies can also prevent discovery of poorly configured applications.
Had the maximum usage policies been in place, we may not have become
aware of the limitations in the FSL deployment: the research team
would have proceeded as before, the cluster would not have been overwhelmed,
the research support group would not have gotten involved, and the
research team may have given up their science aims to work with the
HCP900 data set because they perceived their local resources as inadequate.

\subsection{Learning by Example}

The preparatory work that the Bedpostx application does to manage
the staging of data pipelines provides a useful example of workflow
construction and management. The complexity of managing multiple phases
of a workflow can be hidden from the end user, providing them with
a simple command-line tool that can hide significant processing effort
via automated job submission. In some respects, these wrappers can
be considered early incarnations of domain-specific science gateways,
hiding complex cluster details behind a simplified user interface.
Exposure to such workflow implementations helps expose research support
teams to opportunities for improving the local cluster environment
for users.

The Bedpostx implementation also demonstrates the utility of developing
multi-job workflows with inter-job dependencies. For example, the
Bedpostx workflow for interacting with batch computing includes a
pre-processing step that splits MRI images into chunks, either by
MRI slices for traditional CPU bound jobs or by groups of slices based
on the number of available GPUs. The next step in the workflow is
the actual computation on each chunk. The workflow concludes with
the reassembly of the chucks into a final data set for following stages.
Arranging this workflow as a sequence of job dependencies allows each
stage to request only its required resources. The pre- and post-processing
steps are lightweight, sequential processes that simply split or reassemble
data files. The computational step they surround can then request
the resources it needs. This tunes resource consumption specific to
each stage of the workflow.

\section{Conclusion}

More than anything, this is a record of the depth of customization
necessary to support modern research pipelines. Modern computational,
research effort is dominated by big data sets. Working with these
data sets puts pressure on network, compute, and storage resources.
Identifying and eliminating resource constraints is crucial to facilitating
science at this scale. Network, compute, and storage resource constraints
are often solvable with simple upgrades to hardware that increases
capacity. This paper detailed improvements to network and compute
resources at the campus which directly stemmed from the effort to
process the HCP900 data set. The improvements will help support processing
of terabyte scale data sets. Additional improvements are planned by
connecting the cluster file system to high bandwidth Science DMZ DTNs
and expanding availability of GPU-enabled applications to prepare
for future large scale data sets. 

Software enhancements to support different platform features tend
to have greater expenses and are not always immediately apparent.
The more a local environment diverges from the environment from the
group that develops the tools, the greater likelihood local site maintainers
will need to adapt and migrate those tools to work efficiently on
local platforms. Without a project framework to integrate multi-platform
support from the local site with the upstream code base, these modifications
will be left to individual discovery of each site using the tools.
Depending on the level of maturity in delivering systems that support
specific domain science, the understanding of tools, and the availability
of labor, these problems may or may not be solvable at specific sites.
In order for these enhancements to reach the widest audience, such
enhancements should be developed openly to ensure the maximum reach
of community contributions to those without the means to solve them.

\section{Acknowledgements}

This work was supported in part by the National Science Foundation
under Grant No. OAC-1541310, the University of Alabama at Birmingham,
and the Alabama Innovation Fund. Any opinions, findings, and conclusions
or recommendations expressed in this material are those of the authors
and do not necessarily reflect the views of the National Science Foundation
or the University of Alabama at Birmingham.

Data were provided by the Human Connectome Project, WU-Minn Consortium
(Principal Investigators: David Van Essen and Kamil Ugurbil; 1U54MH091657)
funded by the 16 NIH Institutes and Centers that support the NIH Blueprint
for Neuroscience Research; and by the McDonnell Center for Systems
Neuroscience at Washington University.

\bibliographystyle{plain}
\bibliography{analyzing-hcp900}

\end{document}